# The Photonic TIGER: a multicore fiber-fed spectrograph

Sergio G. Leon-Saval, Christopher H. Betters and Joss Bland-Hawthorn
School of Physics, University of Sydney, NSW 2006, Australia

**ABSTRACT**

We present a proof of concept compact diffraction limited high-resolution fiber-fed spectrograph by using a 2D multicore array input. This high resolution spectrograph is fed by a 2D pseudo-slit, the Photonic TIGER, a hexagonal array of near-diffraction limited single-mode cores. We study the feasibility of this new platform related to the core array separation and rotation with respect to the dispersion axis. A 7 core compact Photonic TIGER fiber-fed spectrograph with a resolving power of around R~31000 and 8 nm bandwidth in the IR centered on 1550 nm is demonstrated. We also describe possible architectures based on this concept for building small scale compact diffraction limited Integral Field Spectrographs (IFS).

**Keywords:** astronomical instrumentation, multicore fiber, spectrograph, photonic lantern, diffraction limited, single-mode fiber.

## 1. INTRODUCTION

Building compact spectrographs at high/medium resolution is always a challenge. Major drivers in increasing astronomical spectrograph sizes, especially in fiber-fed ones, is their multimode nature and the size of optics.

Conventional astronomical fiber-fed spectrographs use multimode optical fibers to feed light from a telescope to the spectrograph slit. These fibers offer a more efficient way of capturing the light imaged at the focal plane of telescopes than their telecom counterpart single-mode fibers [1]. However, the multimode behavior of these spectrographs governs their size and resolution, being far away from the more efficient and compact diffraction limited platforms. In fiber-fed spectrographs it is possible to achieve diffraction limited performance by using single-mode optical fibers. These fibers guide light in a near-diffraction limited Gaussian-like beam, forming the smallest possible input slit for a spectrograph and resulting in cleaner point spread functions (PSF). However, single-mode fibers are not widely used in astronomy due to the low coupling efficiency obtained when coupling light from a telescope. Nowadays, the use of single-mode fibers is possible due to development of the photonic lanterns [2,3,4]. These are efficient multimode to single-mode waveguide convertors that allow the use of diffraction limited spectrographs to be fed by telescopes without the need to match their focal ratios or indeed to worry about the multimode nature of their beam [5,6]. Furthermore, these mode convertors had allowed the use of state of the art single-mode photonics techniques to be implemented in conventional telescope instruments, such as Fiber Bragg Gratings for sky OH suppression [7]. Single-mode and few-mode fibers have also been explored on astronomical instruments with adaptive optics (AO) systems [8], however there has never been a real drive to explore single-mode fiber-fed spectrographs until recently. New efforts are underway to develop a new platform of compact high resolution diffraction limited spectrographs [5,9,10].

The size of optics (hence the spectrograph) could be somehow reduced by using single-mode instead of multimode fibers due to the more forgiving Gaussian PSF produced by those single-mode fibers, thus smaller optics can be used with less significant off-axis aberrations. However, the size of the optics is still very much related to the number of fibers used in the spectrograph. To date fiber-fed astronomical spectrographs are used in a pseudo-slit configuration [11], the fibers are placed in a 1D linear array at the entrance of a classical long slit spectrograph. The image of the dispersed spectra onto the detector is a 2D dimensional representation of the data; one spatial, corresponding to the individual fiber spectra, and one spectral, corresponding to the disperse spectrum of each individual fiber. The width (i.e. the number of fibers) of these fiber pseudo-slits is indeed another important limiting factor in the compactness of the spectrograph. Due to off-axis aberrations, particularly at high resolution, the size of lenses and dispersion optics had to be many times larger than the pseudo-slit size in order to avoid them. The compactness and throughput of high resolution spectrographs that use large number of fibers, such as the case of Integral Field Spectroscopy (IFS) and Multi-Object Spectroscopy (MOS) are often compromised; needing extremely large spectrographs for optimum performance.

In this paper we describe a new diffraction limited fiber-fed spectrograph architecture based in a 2D pseudo-slit formed by a hexagonal array of single-mode cores. A slight rotation between the dispersion axis of the spectrograph and the multicore array avoids spectra overlapping on the detector. This dispersion geometry onto a detector was used for the first time in a bulk optical astronomical instrument called TIGER [12,13], which used microlenses array to perform bidimensional high spatial resolution spectrographic observations. In recognition to this, we decided to call our instrument the Photonic TIGER. Our approach is a combination of classical refractive optics and a new fiber bundle pseudo-slit concept.

## 2. PHOTONIC TIGER MULTICORE ARRAY INPUT GEOMETRY

### 2.1 Pseudo-slit size: conventional versus Photonic TIGER

In conventional pseudo-slits, fibers are stacked vertically to form a long slit [11]. For large slits, they are placed on a radius whose center of curvature coincides with that of the collimator. Additionally, fibers are often aimed in a fanlike pattern outward from the center curvature toward the collimator, so that the central (gut) ray from each fiber strikes the collimator normal to the surface. All those cumbersome approaches are indeed necessary to reduce off-axis aberrations and the size of optics (hence the spectrograph). In a conventional pseudo-slit the width of the slit, i.e. diameter of the lens needed, increases proportionally with the number of fibers. This obviously will depend on the aberration tolerances and resolution of the spectrograph; however this is often the case. In the other hand, the Photonic TIGER pseudo-slit can be built by a multicore fiber with a hexagonal array of cores or by a hexagonal fiber bundle. This geometry allows increasing the number of fibers in the entrance slit while keeping the overall main off-axis distance of all the fibers to a minimum (see Fig. 1), thus the size of the optics in the system.

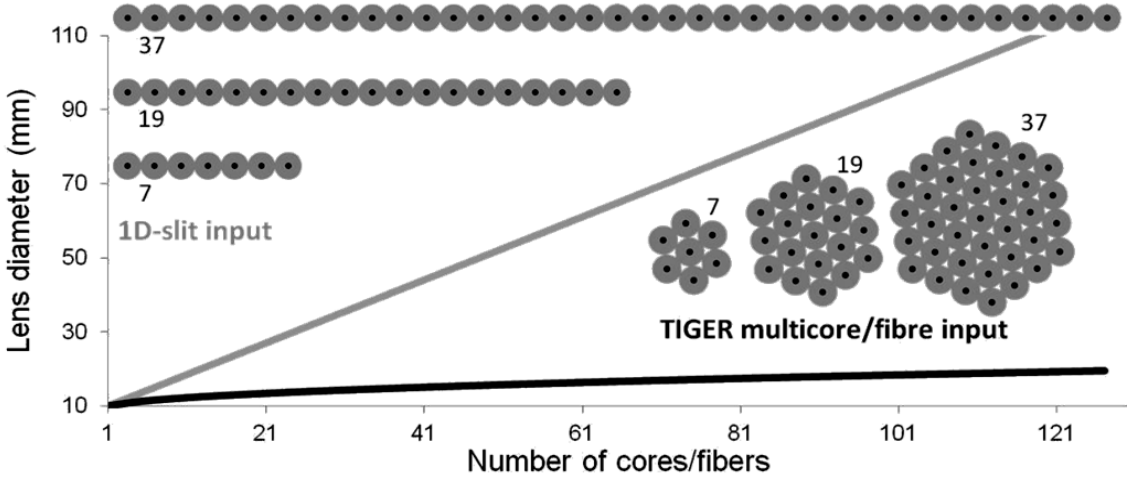

Fig. 1. Spectrograph' lens diameter comparison between a conventional and a Photonic TIGER pseudo-slit versus the number of cores/fibers in the spectrograph.

It is obvious from Fig. 1. that the Photonic TIGER offers size advantages over the conventional pseudo-slit geometry. The results shown are just an approximation based in an extremely compact diffraction limited moderate resolution spectrograph, R~1000 for satellite applications reported in [9]. The single-mode cores are separated by their fiber size, 127 μm, and designed to work in the visible region 450-700 nm with Numerical Apertures (NA) of 0.1. We used ZEMAX to calculate the smallest lens diameter possible to achieve near-diffraction limited performance from all the cores/fibers in the slit using a 12.7 mm effective focal length in our system. A 37 fiber conventional pseudo-slit will have a slit width of ~4.6 mm, based in our calculations (see Fig. 1) this slit will need a collimator lens diameter of ~40 mm. In the other hand, a 37 core/fiber Photonic TIGER slit will have a ~0.77 mm effective slit width needing a much smaller ~15 mm diameter lens. This is only an approximation to picture the idea behind the concept, a more detail study for accurate values will have to be done, and this should include several different focal lengths, which indeed should affect these results.

## 2.2 Hexagonal core array parameters

Other important factor to consider is the rotation and the core separation of the hexagonal array to avoid overlapping spectra onto the detector. In order to produce $N_c$ non-overlapping spectra, each corresponding to an individual core of the array, the cores must have a sufficient separation perpendicular to the dispersion axis on the detector. Given that the PSF of diffraction limited spectrographs should always be Gaussian in nature, the cores vertical separation should be at least the $1/e^2$ width of the spectra on the detector. For simplicity we will assume here that the spectrograph produces a 1:1 image on the detector, thus the fiber separation could be measured in multiples of the core's mode field diameters (MFD). We need to determine the optimal separation and core rotation angle, $\theta$, of the multicore fiber. A schematic diagram of the hexagonal grid of 7 cores that has been rotated by an angle $\theta$ is given in Fig. 1(a). In order to evaluate the ideal geometry we start by equating the vertical separation ($v_{sep}$) of adjacent cores (i.e A and O) and cores at opposite edges of the middle row and middle ±1 row (i.e. B and F). In doing so the rotation required to maximize the separation of each core in the fiber is found.

The vertical separation of adjacent cores A and O is simply,

$$v_{sep} = a = u \sin \theta \tag{1}$$

To find the separation of B and F, labeled $a$ in Fig. 2(a), we first note that the angle $\angle CBD$ is also $\theta$. We can then see from Fig. 2(a) that

$$a = (d + l)\cos\theta - b \tag{2}$$

Where $d$, $l$ and $b$ are,

$$d = u\sqrt{3}/2, \quad l = \tan\theta\, u/2, \quad b = N_{mid}\, u \sin\theta \tag{3a,b,c}$$

Eqns. (3a,b,c) are a consequence of the regular spacing ($u$) of the hexagonal grid. In the 7 core case shown in Fig. 2(a), $N_{mid}$ is 3, but this result is general for any size array. By equating Eqns. 1 and 2 we get an expression for the optimal rotation angle for a hexagonal core array with $N_{mid}$ cores in the middle row or $N_c$ cores in total, Fig. 2(b) shows a plot of the optimal angle of rotation versus the number of cores.

$$\tan\theta = \frac{\sqrt{3}}{2N_{mid}-1} = \frac{3\sqrt{3}}{2\sqrt{12N_c-3}-3} \tag{4}$$

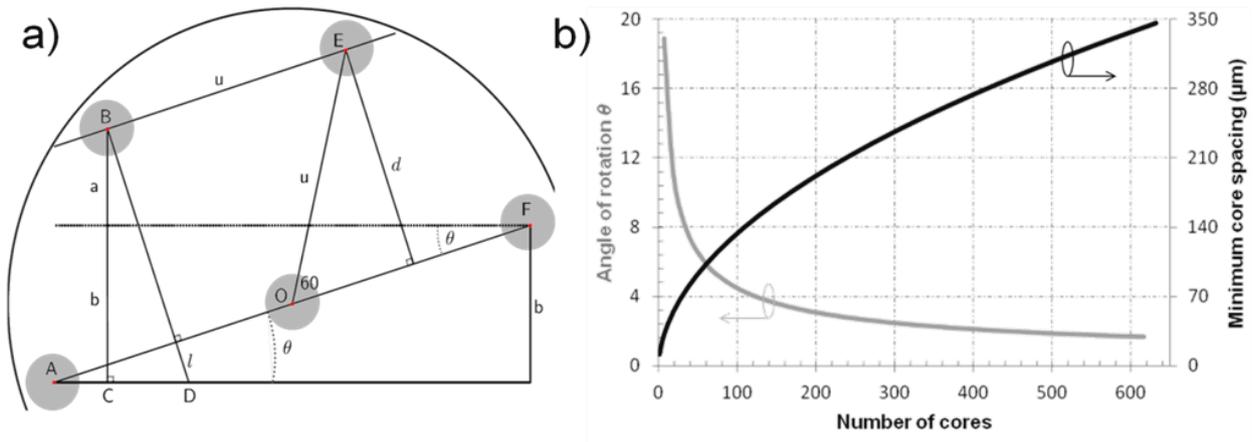

Fig. 2. (a) Schematic diagram of the hexagonal array core spacing and geometry in a 7 core array (only 5 cores shown). The cores are spaced by a distance $u$ and the array is rotated such that each row of cores is at an angle $\theta$ to the horizontal axis. (b) Angle of rotation and minimum optimal core spacing versus the total number of cores. This is calculated by considering a 1:1 imaging and an individual core MFD of 10.5 µm.

Eqn. 4 shows that the rotation required to optimize the vertical separation of a given hexagonal array size, is in principle independent of the core spacing, however this alone does not guarantee well separated spectra on the detector. By eliminating $\theta$ in Eqn. 4 we get an expression describing the necessary core spacing to achieve a given or indeed desired vertical core separation $v_{sep}$,

$$u = \frac{2\sqrt{(1-N_{mid}+N_{mid}^2)v_{sep}^2}}{\sqrt{3}} \quad (5)$$

Fig. 2(b) shows a plot of the optimal core separation versus the number of cores assuming a MFD of 10.5 µm for each individual core and a 1:1 imaging setup on the spectrograph. The 10.5 µm MFD is the standard value for a single-mode fiber core at 1550 nm wavelength. However these calculations will be valid for any other wavelength or different MFD cores, in that case a simple change of the $v_{sep}$ in Eqn. (5) will give us the minimum core spacing for the desired array parameters.

We used the calculated values for the optimal rotation and core separations in order to simulate the dispersed spectra images onto the detector. A range of possible Photonic TIGER pseudo-slits were considered from 7 to 127 cores. The results are shown in Fig. 3.

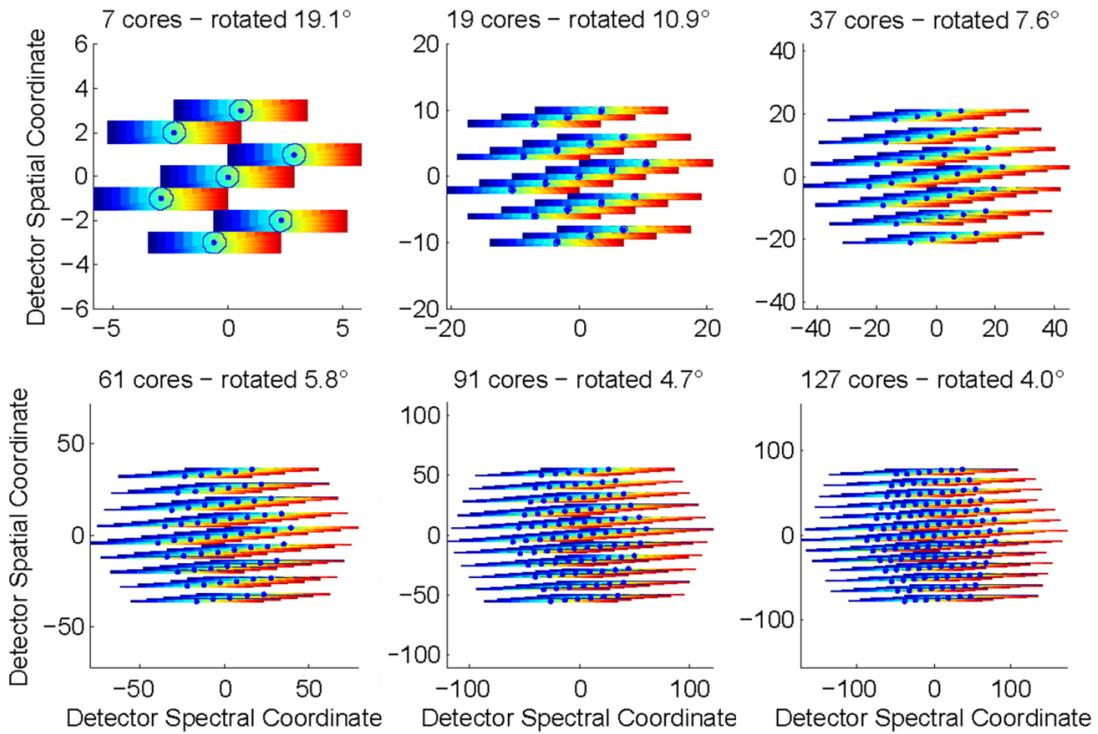

Fig. 3. Simulated spectra with a vertical separation equal to the 1/e² width of a single spectrum marked by a blue circle around each core. Six different core arrays 7, 19, 37, 61, 91 and 127 with their respective optimal angle rotation are shown. The wavelength range shown in each is set to twice the distance in the spectra coordinate of the edge core in the middle row from the origin. The wavelength increase on the detector goes left to right, as indicated by the colors (blue to red).

In the Photonic TIGER architecture the reduction in the spectrograph size comes with a cost. As it can be seen on the simulated spectra, as it is also the case for many classical multi-slit spectrographs, some of the spectra could be truncated at the edge of the detector. The spectral length in pixels must be chosen and limited to a fraction of the detector linear size to avoid too large a fraction of truncated spectra. In particular the far left cores imaged on the detector will limit the common bandwidth for all the fibers in the blue span of the spectra and the far right ones the red part.

# 3. EXPERIMENTAL PHOTONIC TIGER FIBER-FED SPECTROGRAPH

## 3.1 IR diffraction limited spectrograph

The Photonic TIGER spectrograph is based on a compact diffraction limited IR design. The demonstration, design, and performance of this prototype spectrograph are reported in Betters et al [10]. Its current configuration has a bandwidth of 8 nm centered on 1550 nm with a resolving power, $\lambda/\Delta\lambda$, of 31000. The spectrograph is based on 1120 l/mm Volume Phase Holographic (VPH) grating, and uses a combination of 1 and 2 inches diameter lenses. The spectrograph has a 70% throughput (light from the single-mode entrance slit that lands on the detector), achieving 85% of the theoretical limit for Gaussian illumination of a diffraction grating. The spectrograph is also extremely compact with a footprint of just 450 mm x 190 mm as shown in Fig. 4, and it is also built with readily available off-the-shelf parts and optics.

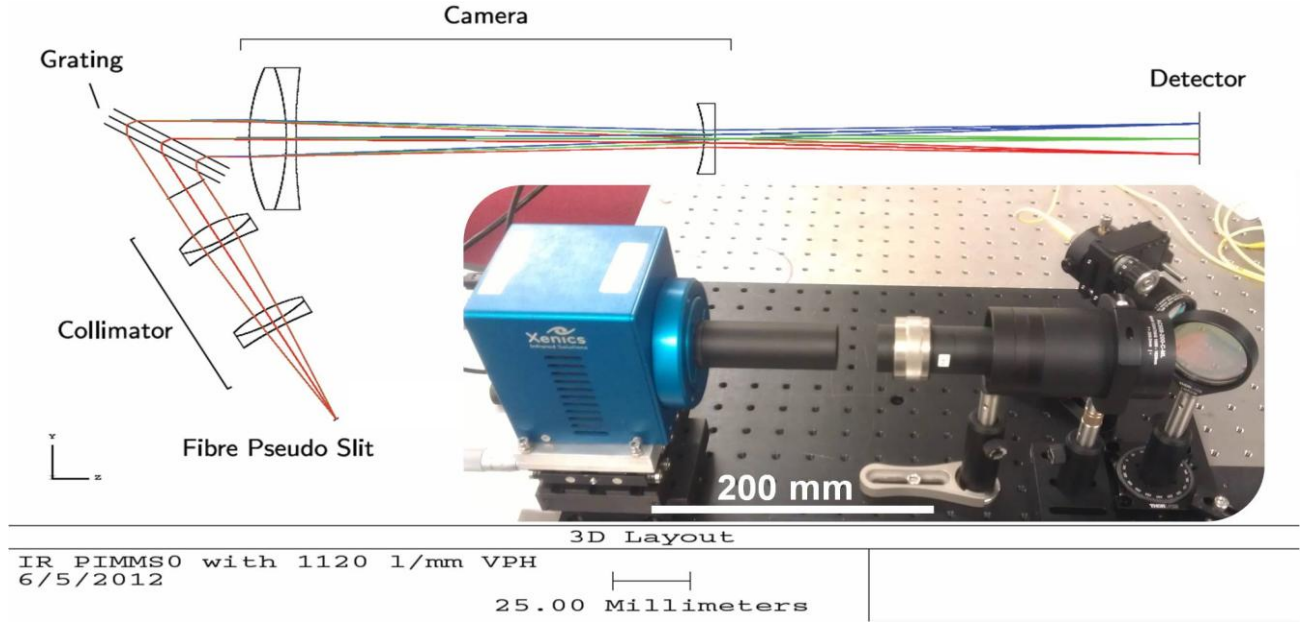

Fig. 4. Optical layout of the IR diffraction limited spectrograph used for the demonstration of the Photonics TIGER. The wavelengths shown are 1550 ±4 nm. (insert) Optical photograph of the IR spectrograph which design fits in almost half of a 450 mm by 300 mm optical breadboard.

The bandwidth and efficiency of the spectrograph is limited by the IR detector. We used a very simple detector, the Xenics[d] Xeva 1.7 320 InGaAs CCD camera. This has a small array, 320x240, of large 30 μm pixels which are sensitive to the NIR (1 to 1.7 μm).

## 3.2 Multicore/fiber-bundle fabrication

In order to proof the concept of the 2D Photonic TIGER pseudo-slit we fabricated a 7 core prototype using our in-house glass processing facility. The fabrication process is similar in nature to that of other devices developed in the area of Astrophotonics [14], such as the photonic lanterns [3,4] and hexabundles [15]. The single-mode fibers used for this purpose were standard silica SMF-28 from Corning, USA, designed for telecom. The 7 fibers were bundled together inside a silica capillary as shown in Fig. 5 (a) left, and then they were fused together over a 2 cm length in a slow process while maintaining the initial cores size. Fig. 5 (a) center and right shows the half way fused and the fully fused 7 core Photonic TIGER pseudo-slit respectively. Once the fusing process was finished, we connectorized (glued and polished) the device with a standard SMA 905 bare fiber connector that can be easily mounted into our IR spectrograph collimator input (see Fig. 5(b)). Our prototype was a seven fiber pigtailed Photonic TIGER pseudo-slit with 7 cores separated by 122 μm with MFDs of 10.5 μm at 1550 nm. The overall diameter of the bare fully fused device was 490 μm.

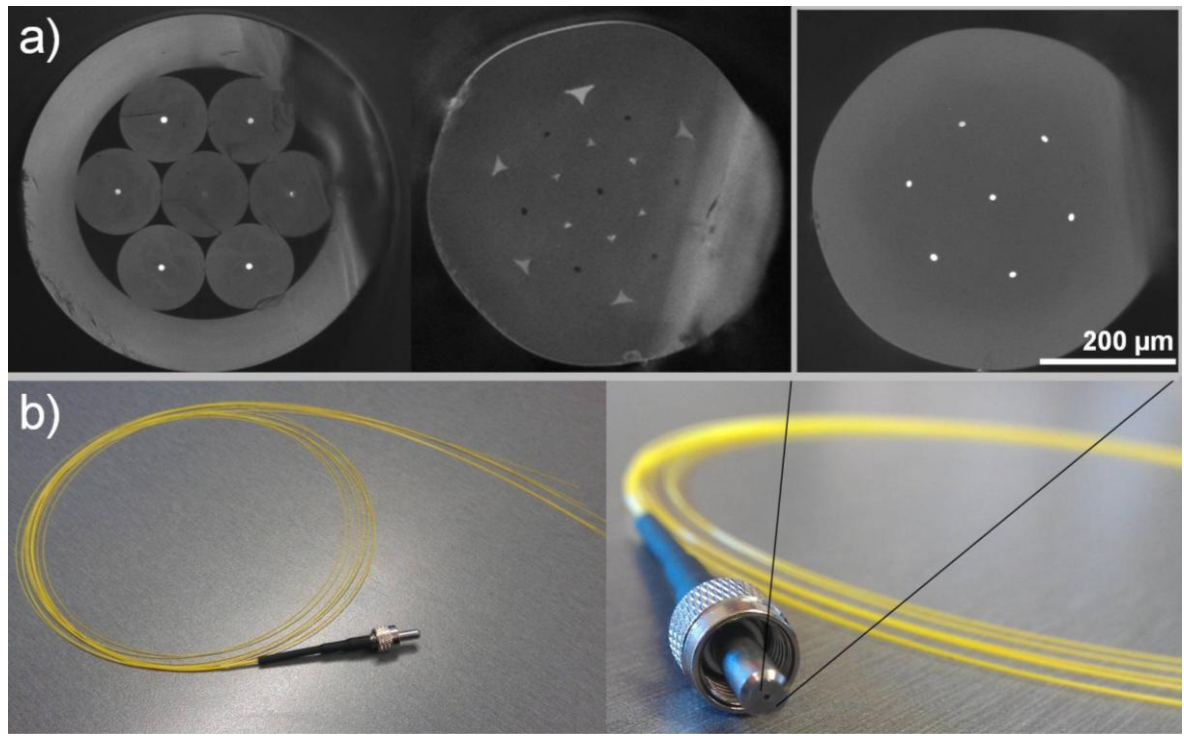

Fig. 5. Optical photographs of (a) 7 core Photonic TIGER pseudo-slit in three different stages of the fusing, packaged, half-way fused and fully fused (left, center and right respectively); (b) Fully fused device connectorized for building the Photonic TIGER spectrograph.

### 3.3 Photonic TIGER spectrograph

We used two different laser sources and a wavelength meter to characterize and calibrate our spectrograph. To be able to couple light to the 7 cores equally, two Photonic TIGER pseudo-slits were spliced back to back, i.e. we spliced the 14 (2x7) single-mode fibers together and use one of the 7 core fused fibers as the light input for our fiber-fed spectrograph. As an anecdote, we would like to note that to our knowledge this setup was indeed the very first single-mode fiber Integral Field Unit (IFU) and hence probably the first true diffraction limited fiber-fed IFS. We collimated the output of a standard single-mode fiber and used that to couple light into our IFU, hence exciting all the cores more or less equally. The single port of a telecom 1x4 single-mode fiber splitter was used in reversed order as the collimated single-mode fiber beam. These combiners are not very efficient, however in our case we used it to couple different laser sources into our spectrograph at the same time for characterization. The use of the two different laser sources gave us the chance to image the cores and the disperse spectra at the same time on the detector. To image a single spot for each core on the detector after the dispersion optics we used a 2 pm narrow bandwidth PHOTONETICS TUNICS-PRI tunable laser source 1520-1600 nm. We simultaneously coupled a 90 nm broadband Thorlabs SLD source centered on 1550 nm. This gave us dispersed spectra onto the detector. The clear image of the cores and the spectra on the detector helped us to evaluate the Photonic TIGER pseudo-slit rotation against the dispersion axis of the spectrograph as shown in Fig. 6. Thanks to the SMA connectorized input of the spectrograph and a standard Thorlabs rotation and XY adjustment stage it was trivial to rotate and fix the orientation of the core array.

From our calculations in Section 2.2 a 7 core Photonic TIGER pseudo-slit will need a minimum core separation of 33 µm and a rotation angle of 19.1°. Fig. 6 shows the raw (without background or dark frame subtraction) detector image of our spectrograph at 4 different rotation configurations 0°, 8.8°, 12.5° and 25.5°. Our prototype was not optimized, having a much larger 122 µm core separation; hence the span of possible rotations for avoiding spectra overlap was very large. The spectra showed no overlapping for rotations between 9.5º and 35.5º.

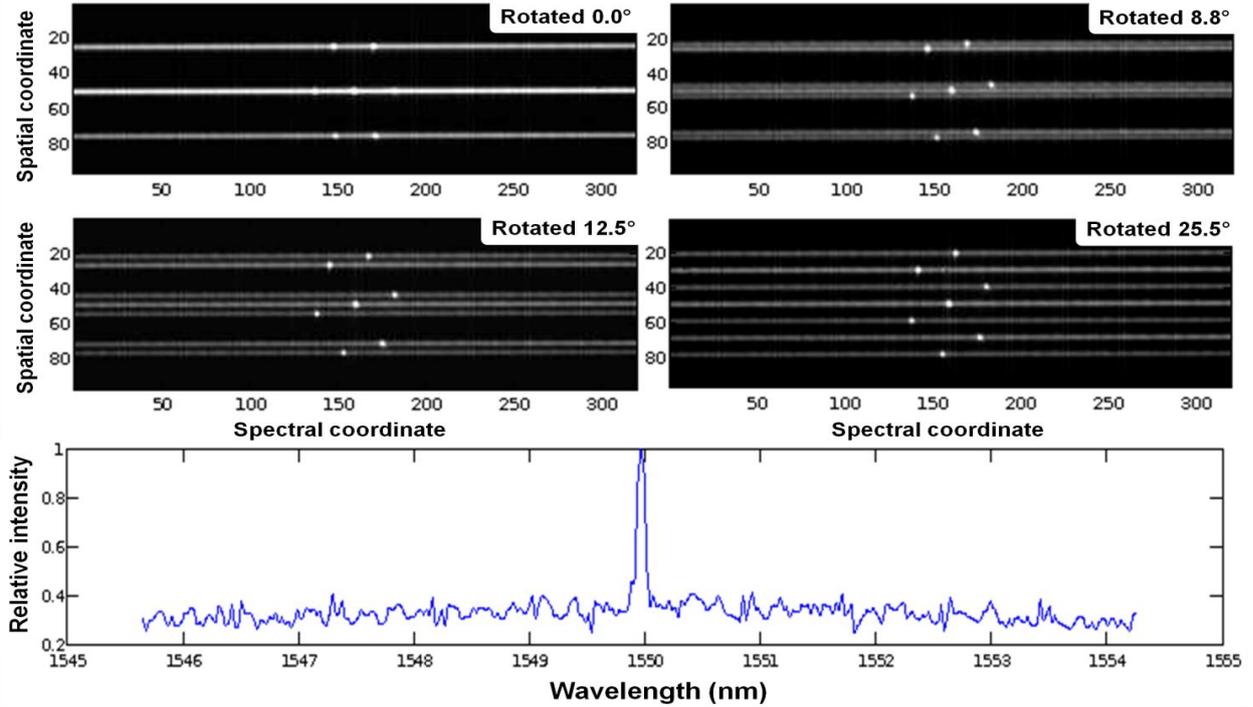

Fig. 6. (four top panels) Spatial versus spectral coordinates detector image of the dual laser source fed Photonic TIGER spectrograph for 4 different pseudo-slit angle rotations 0°, 8.8°, 12.5° and 25.5°, images are 100 x 320 pixels, raw data without background or dark frame subtraction. (bottom panel) 1D processed spectrum from the central core in the 25.5° rotated case showing the broadband source spectrum and the overlapped narrow laser source tuned at 1549.95 nm.

The 2 pm narrow bandwidth of our laser could be resolved to an R~31000. The calibration of the spectrograph was done by using a combination of the narrow laser source and a Burleigh WAVEMETER WA-1100 with an absolute wavelength accuracy of ± 2.5 pm. The TUNIC tunable laser was coupled to the spectrograph at the same time that a second output of the 1x4 single-mode fiber combiner was monitored with the wavelength meter. The high resolution of the spectrograph also showed the multimode emission effect on the spectrum of our broadband laser diode source. This can be observed in the fine ripple structure of the 1D processed spectrum of one of the single cores, showed in the bottom panel of Fig. 6.

### 3.4 Photonic Tiger spectrograph configurations and future work

The Photonic TIGER spectrograph can be implemented at any wavelength and in theory with any (feasible) resolution. In its current configuration the spectrograph has limited applications, due to the difficulty of coupling light from a telescope into a single-mode core [1]. However, there have been some interest in using single-mode and very few-mode fibers in telescopes with AO systems and extremely high resolution R~200000 have been demonstrated with this configuration [8].

A more powerful concept for this platform is the use of photonic lanterns [5] in combination with larger core number Photonic TIGER. A multimode input converted to single-mode outputs means no need for complicated AO systems and cumbersome optics to achieve near-diffraction limited performance spectroscopy in almost any telescope. These spectrographs will accept light from (almost) any input f/ratio, either natural seeing or AO-corrected, any input fiber diameter, and any telescope diameter. To date photonic lanterns have been demonstrated in three different platforms; single-mode fiber pigtailed [3,4], multicore fiber [16], and bulk glass ultrafast laser inscription waveguides [17]. Our next prototype will consider fiber pigtailed lanterns and multicore fiber lanterns. Fiber pigtailed Photonic TIGER pseudo-slits with 61 and 91 cores will be soon fabricated and implemented in a visible Photonic TIGER spectrograph in our facilities. A most exciting alternative to this concept will be multicore fibers with the right number of cores and geometry, this alternative is already under consideration. Furthermore, fabricating photonic lanterns from those multicore fibers will reduce to an absolute minimum the total number of fibers in the system. The final approach will be to combine hexabundles, multicore photonic lanterns and 2D multicore fiber pseudo-slits into the same spectrograph: the Photonic TIGER Integral Field Spectrograph.

One of our aims for this platform is also to create a range of extremely light, compact and low cost spectrographs with medium and low resolution for space based applications, such as pico-, nano-satellites and balloons. Our current efforts towards this are currently reported in Betters et al [9]. Another interesting area for this type of compact, medium cost spectrographs is in instrumentation for small to medium size telescopes (<2 m).

## 4. CONCLUSION

We have shown a new platform for spectrographs' fiber pseudo-slits that can minimize the size of spectrographs just by using a 2D core/fiber geometry. In doing so we have also demonstrated a powerful R~31000 compact diffraction limited multiple single-mode fiber input fiber-fed spectrograph. This architecture in combination with other astrophotonics technologies, such as photonic lanterns and hexabundles, could represent the first astronomical high spatial and spectral resolution near-diffraction limited multimode fiber-fed IFS.